# Do cosmic rays drive jets?

**AR Bell**

Blackett Laboratory, Imperial College, London SW7 2AZ, UK

E-mail: t.bell@imperial.ac.uk

**Abstract**. A sudden release of high energy cosmic rays at the centre of a wind sustaining a spiral magnetic field produces cavities of low density and low magnetic field along the axis. The trajectories of high energy cosmic rays are focussed onto the axis, and lower energy cosmic rays and thermal plasma can escape through the cavities. This may explain the jets often seen in accretion systems and elsewhere.

The formation of astrophysical jets and the origin of ultra-high energy cosmic rays are two outstanding problems in astrophysics. Here, they are brought together in a model in which cosmic rays generated at the centre of an accretion system initiate jet formation through their interaction with a magnetized wind above the accretion disk. Supersonic motion and shocks accompanying accretion are a natural source of cosmic rays. Low energy cosmic rays are initially confined near the shocks, but ultra-high energy cosmic rays, escaping into the magnetized wind, push aside the plasma to form cavities of low density and low magnetic field on the rotation axis. The trajectories of the highest energy cosmic rays are focussed onto the axis, and lower energy cosmic rays and thermal plasma may then escape through the cavities to form relativistic jets.

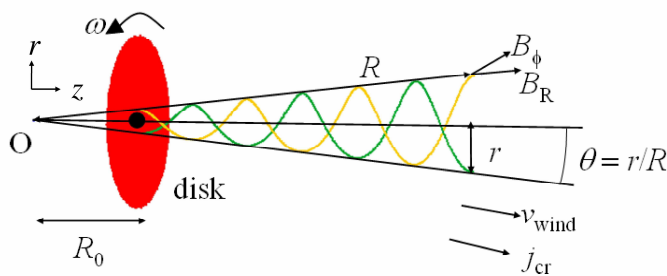

**Figure 1**. Idealised model of a helical magnetic field in the wind from an accretion disk.

An idealised model of a wind formed by a rotating accretion disk is shown in figure 1. The wind to the right of the disk propagates as though flowing radially away from the point O which is a distance $R_0$ behind the disk. The disk rotates with an angular velocity $\omega=\omega_0(r/R_0)^{-3/2}$ as appropriate for Keplerian rotation, where $\omega_0$ is a constant. The rotation of the disk produces a spiral magnetic field pattern in the wind [1]. The components of the magnetic field are $B_R=B_{R0} R_0^2/R^2$ and $B_\phi =(\omega_0 R_0 B_{R0}/v_{wind})$ $(z/R)$ $(R_0/R)$ $(z/r)^{1/2}$ in cylindrical co-ordinates $(r,z,\phi)$ where $R=(r^2+z^2)^{1/2}$ and $B_R$ is the component of magnetic field in the $R$ direction. $B_{R0}$ is the magnetic field normal to the disk at its centre. The ratio of $B_\phi$ to $B_R$ is determined by the ratio of the disk rotation velocity $\omega r$ to the wind speed $v_{wind}$ since the magnetic field is frozen into both wind and disk. We assume that $v_{wind}$ is independent of position. The following analysis could be adapted to more complicated configurations.

Suppose that a sudden energy release occurs near the centre of the accretion disk and that much of this energy is converted into cosmic ray (CR) particles as suggested, for example, in [2]. The maximum CR energy attainable by diffusive shock acceleration is that at which the CR mean free path, and the CR Larmor radius to which it is approximately equal [3], become sufficiently large for the CR to escape the acceleration region into the wind. The efficiency of diffusive shock acceleration of CR is probably in the range 10-50%, and CR accelerated by non-relativistic shocks have equal energy densities in each decade of energy (spectra can be even flatter for relativistic shocks), so the high energy CR flux should carry a substantial fraction of the available energy. If the CR particles are protons or positive ions, they carry a current density $j_{CR}$ and are subject to a $j_{CR} \times B$ force as they cross the magnetic field in the wind. If the component of the field normal to the disk surface is anti-parallel to the rotation vector, the CR trajectories are focused towards the axis. Momentum conservation requires that the background wind plasma is correspondingly pushed away from the axis. If the magnetic field and the rotation vector are parallel rather than anti-parallel, the CR trajectories are instead deflected away from the axis and background plasma pushed toward the axis. The force on the background plasma is mediated by $j_t \times B$, where the current $j_t$ carried by the background thermal plasma is opposite to the CR current ($j_t \cong -j_{CR}$) as required to maintain charge neutrality in the wind [3,4]. If the CR are focused towards the axis on one side of the disk, the same should apply on the other side.

The geometry close to the axis is similar to that considered in detail in section 2.2 of [4] in which a uniform current density $j_{CR}$ of CR is parallel to the axis of a helical magnetic field. It was shown that, in the anti-parallel configuration, if the $j_{CR} \times B$ force dominates all other forces, the radius $r$ (defined in figure 1) of a particular helical magnetic field line increases exponentially in time with a growth rate $\gamma=(j_{CR}B_\phi/\rho r)^{1/2}$ where $B_\phi$ is the local azimuthal magnetic field, $\rho$ is the local wind density, and it is assumed that the current density $j_{CR}$ is unaffected by change in the wind structure. The growth rate for each field line is constant since $B_\phi/\rho r$ is constant on each field line as it expands in cylindrical geometry. Because the azimuthal field is larger close to the axis, the growth rate there is also larger, and the inner field lines expand more rapidly. They cannot overtake the outer field lines, so the growth is not simply exponential, and a cylindrical snow-plough is formed which approaches self-similar expansion with a radius $r_s$ proportional to $t^{4/3}$. Close to the axis, all quantities vary more slowly in $z$ than in $r$ so the $z$ dependence can be neglected and the evolution of the snow-plough can be represented by the one-dimensional (in $r$) equations 7 in [4]. The equations can be solved to find the profile as a function of radius $r$ across a cross-section of the snow-plough. Figure 2 shows the results of time-dependent numerical solution as the profiles approach self-similarity in a case in which the magnetic pressure is not significant and the $j_{CR} \times B$ force dominates. The velocity $u_r$ is normalised to $r_s/t$. The magnetic field $B_\phi$ and density $\rho$ are normalised to their initial values at radius $r_s$.

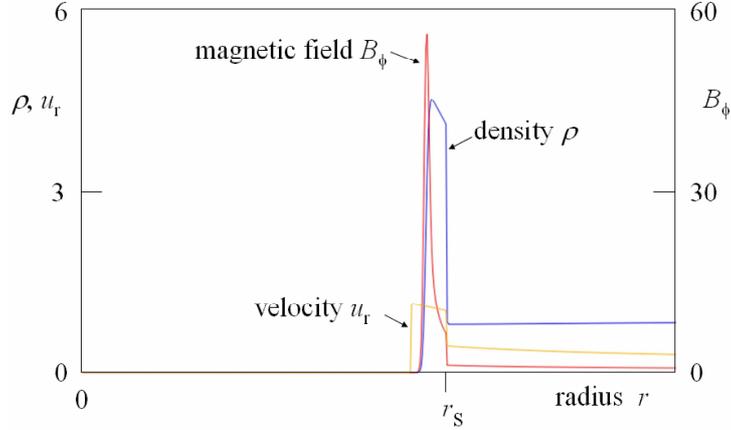

**Figure 2**. Profiles of radial velocity $u_r$, density $\rho$ and azimuthal magnetic field $B_\phi$.

Inside the snow-plough, a cavity is formed in which both the density and the magnetic field are negligibly small, and the velocity is hence undefined. The magnetic field in the walls of the cavity and the surrounding plasma focus the trajectories of high energy CR into the cavity, and the decrease in magnetic field on axis allows lower energy CR to escape through the cavity. The decrease in mass density implies that any shock expanding into the wind will have a higher velocity along the axis, thus allowing thermal plasma to escape preferentially through the cavity. This suggests that a jet of low as well as high energy CR, and also of thermal plasma, may form along the axis, and, if so, from the symmetry of the magnetic helix, a similar jet will form on the other side of the disk. This provides a natural way for jets to form wherever there is a sudden release of CR into a wind permeated by a spiral magnetic field.

Begelman, Blandford and Rees [5] deduced a set of 'fiducial' parameters characterising the environment of a black hole. The fiducial magnetic field $B_b$ is given by $cB_b^2/2\mu_0 = L_E/4\pi R_b^2$, where $L_E$ is the Eddington luminosity and $R_b$ is the black hole radius. The magnetic field can be generated by shear in fluid motions around the black hole, or by the CR acceleration process itself [3,4,6,7]. A fiducial CR proton energy $\varepsilon$ can further be obtained by equating the Larmor radius in the fiducial field $B_b$ to the black hole radius, giving $\varepsilon = 3 \times 10^{20} M_8^{1/2}$ eV, where $M_8$ is the black hole mass in units of $10^8$ solar masses. CR may be shock-accelerated to this energy, at which their Larmor radius is large enough for them to escape into the wind and produce the cavity. Equivalently, the same estimate for the CR energy, within a numerical factor of order unity, can be obtained by setting the power of the CR proton beam to the Eddington luminosity and the electric current of the beam to the Alfven current [8], which is the largest current a charged particle beam can carry through a neutralising background without pinching under its own self-field (see [4] for further discussion). It is notable that for a black hole of $10^8$ solar masses at the centre of an active galaxy the estimated CR energy is similar to the maximum energy of CR arriving at the earth, whereas the CR energy for a solar mass black hole is comparable with that of the highest energy CR thought to be produced within our own galaxy.

The equations for the CR-wind interaction in the snow-plough configuration have a self-similar solution with the radius of the snow-plough given by

$$r_s = \eta R_0 (R/R_0)^{-1/3} (t/t_0)^{4/3}$$

where $t_0 = (\rho_0 v_{wind}/j_0 \omega_0 B_{R0})^{1/2}$, $\eta$ is a numerical constant and $\rho_0$ and $j_0$ are the wind density and CR current density immediately above the disk in the idealised model of figure 1. The numerical simulation of figure 2 gives $\eta = 1.8$, and this agrees well with a simple analytic momentum- and energy-conserving model for a snow-plough consisting of a thin shell of swept-up mass. At any given time, the cavity appears as an inverted funnel since the radius of the cavity reduces with distance $R$ from the accretion disk, and this may guide CR and thermal plasma into the cavity. The equation for $r_s$ can be re-arranged to give the opening half-angle $\theta_s = r_s/R$ of the cavity:

$$\theta_s = 2(ct/R)^{4/3}(v_{\text{wind}}/c)^{-2/3}(R_b/r_{g0})^{2/3}(R_0/R_b)^{1/3}(I_0/\rho_0 c^3)^{2/3}$$

where $r_{g0}$ is the Larmor radius in the disk field $B_{R0}$ of CR with characteristic energy $\varepsilon$, $I_0$ ($I_0 = j_0\varepsilon$, with $\varepsilon$ in eV) is the CR energy flux at the disk surface, and $R_b$ is the black hole radius defined as $R_b = \omega_0^2 R_0^3 c^{-2}$ at which the rotation velocity is notionally equal to $c$. The magnitude of most terms in the equation for $\theta_s$ may well be of order one, except that $I_0/\rho_0 c^3$, the efficiency with which the high energy CR are produced, is most likely in the range 1-10% [3]. This is consistent with an opening half-angle $\theta_s$ of a few degrees as seen in well-collimated jets. However, the equation for $\theta_s$ only applies to the initial stages of jet formation, and a more complicated model is needed once high energy CR trajectories are focussed onto the axis and lower energy CR and thermal plasma escape through the cavity.

The model presented here is idealised, but encourages the suggestion that the release of CR into a magnetized wind might at least seed the production of the energetic and sometimes very narrow jets observed in accretion systems. Not only can the interaction of CR with magnetized plasma generate the large magnetic field needed to explain galactic CR acceleration to the spectral knee at $10^{15}$eV and possibly beyond [3,4,6,7], but it may also explain the widespread production of astrophysical jets. The general condition for cavity and jet formation is that the growth rate $\gamma$ is greatest close to the axis; this can be true for magnetic field, wind density and wind velocity varying in other ways over the disk surface, as for example for the dependencies assumed in [9]. A similar model might apply to a supernova occurring in a circumstellar wind, although the condition of largest growth rate on axis would be less easily met and jets correspondingly less tightly collimated. Similar processes may also occur for steady or repeated release of CR into a magnetized wind.

**Acknowledgements**
I wish to thank the organisers for the invitation to their very stimulating workshop on *Physics at the End of the Galactic Cosmic Ray Spectrum*.